\UseRawInputEncoding
\documentclass[aps,floatfix,twocolumn,balancelastpage,amsmath,amssymb,longbibliography, 10pt]{revtex4-2}

\usepackage{graphicx}
\usepackage{amsmath}
\usepackage[normalem]{ulem}
\usepackage{comment}
\usepackage{physics}

\newcommand{\yb}{$^{171}\textrm{Yb}$ }
\newcommand{\ybion}{$^{171}\textrm{Yb}^{+}$ }

\newcommand{\mic}{\,$\mu$m\:}

\begin{document}

\preprint{APS/123-QED}

\title{Autonomous multi-ion optical clock with on-chip integrated photonic light delivery}

\author{
Tharon~D.~Morrison$^{1,*, \dagger}$, Joonhyuk~Kwon$^{1,*, \ddagger}$, Matthew~A.~Delaney$^{1}$, Michael~Gehl$^{1}$, David~R.~Leibrandt$^{2}$, Daniel~Stick$^{1,3}$, and Hayden~J.~McGuinness$^{1, \top}$}
\affiliation{
\mbox{$^1$Sandia~National~Laboratories,~Albuquerque,~New Mexico,~87185,~USA} \\ 
\mbox{$^2$Department of Physics and Astronomy, University of California, Los Angeles, California, 90095, USA}\\ 
\mbox{$^3$Current affiliation : IonQ Inc., College Park, MD, USA}\\
$^*$These authors contributed equally to this work.\\$^\dagger$tdmorr@sandia.gov;~$^\ddagger$jookwon@sandia.gov;~$^\top$hmcgui@sandia.gov
}

\date{\today}

\begin{abstract}
 Integrated photonics in trapped-ion systems are critical for the realization of applications such as portable optical atomic clocks and scalable quantum computers. However, system-level integration of all required functionalities remains a key challenge. In this work, we demonstrate an autonomously operating optical clock having a short-term frequency instability of 3.14(5)$\times 10^{-14} / \sqrt{\tau}$ using an ensemble of four \ybion ions trapped in a multi-site surface-electrode trap at room temperature. All clock operations are performed with light delivered via on-chip waveguides. We showcase the system's resilience through sustained, autonomous operation featuring automated ion shuttling and reloading to mitigate ion loss during interleaved clock measurements. This work paves the way beyond component-level functionality to establish a viable and robust architecture for the next generation of portable, multi-ion quantum sensors and computers.
\end{abstract}

\maketitle

\section{Introduction}

\begin{figure*}[!tp]
    \includegraphics[width=0.8\linewidth]{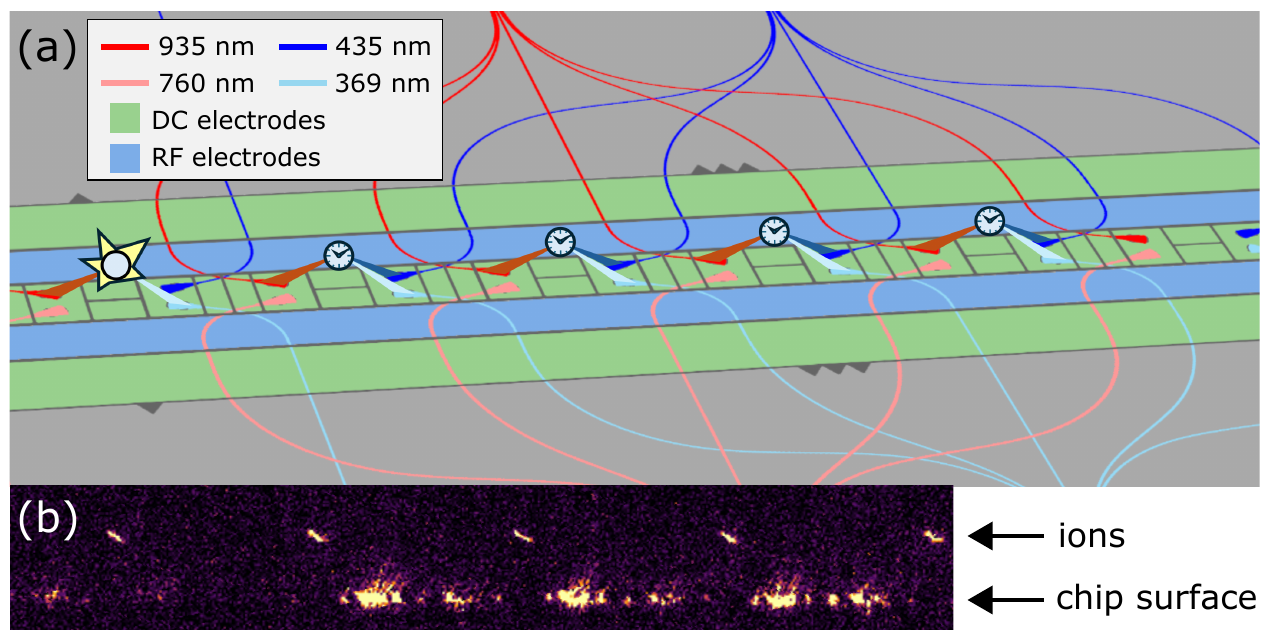}
    \caption{\textbf{Schematic image for the integrated multi-ion trap} \textbf{(a)} Illustration of trapped ions above the surface of the integrated waveguide chip, with the loading site indicated by the star-burst and the clock sites indicated by clock icons. The trapping regions where the ions are addressed consist of waveguides on two different embedded layers (reds and blues), DC electrodes (green) and RF electrodes (teal) both on the top layer. \textbf{(b)} Camera view of trapped ions. As a check, ions are aligned to a camera prior to fine alignment with a multi-channel PMT array. Ions are imaged from the side to reduce background photons from the light scattering out of the waveguides and off the chip surface. }
    \label{fig:SystemIntro}
\end{figure*}

Optical atomic clocks, as time and frequency standards, offer unparalleled precision for applications in fundamental physics \cite{RevModPhys87637,PhysRevLett116063001,Safronova:2018,brewer:2019a,hb3cdk28,PhysRevLett134093201}, navigation \cite{Burt:2021}, and telecommunications \cite{Lindvall:25,sheridan2025telecommunications}. 
Neutral atom optical lattice clocks are widely used platforms that provide high precision and accuracy \cite{Takamoto:2005,Boyd:2005,Bothwell:2019,McGrew:2018} but contain complex atom-atom interactions and require large high-power laser systems and atomic ovens. On the other hand, trapped-ion clocks \cite{Chou:2010, hausser2025in, brewer:2019a}, provide excellent isolation of individual ions with minimal disturbance of internal states. However, they suffer from high quantum projection noise as a result of the comparatively small number of ions. Although both platforms serve as excellent clocks, they both rely on bulky optical components, which limit their operation to laboratory environments.

The path to deploying these instruments in real-world applications lies in miniaturization and ruggedization \cite{Newman:2018,Burt:2021}, a challenge for which integrated photonics provides a compelling solution. By replicating complex optical functionalities on scalable chip-based platforms \cite{chiaverini:2005, stick:2005}, this approach not only works to reduce the system's footprint but also enables advanced multi-ion clock architectures that are less practical with conventional setups \cite{rosenband:2013,blain:2021}. Additionally, the integration of photonic components directly into ion trap chips is a key strategy for scaling these quantum systems. Replacing complex free-space optics with on-chip waveguides and splitters \cite{sundaram2022highly, ricci_standingwave_2023, Kwon:2023}, grating couplers \cite{Mehta2016,niffenegger:2020,mehta:2020}, and photonic control \cite{hogle:2023} provides robust and scalable delivery of the multiple laser wavelengths required for narrow transition control. These technologies are also applicable to trapped-ion quantum computing, notably the Quantum Charge-Coupled Device (QCCD) architecture \cite{noel:2019, pino:2021}, which requires multiplexed operations and coherent control across many distinct trapping zones \cite{Mordini:2025}. By providing a pathway to individually address dense ion arrays, integrated photonics are critical for moving beyond small-scale experiments, towards large-scale quantum platforms.

Recently demonstrated building blocks for this architecture include simultaneous multi-site optical addressing \cite{Kwon:2023} and coherent transport of ions between zones \cite{Mordini:2025}, establishing the viability of complex on-chip control primitives. Although these advances are foundational, they have so far focused on individual capabilities rather than their combination into a unified, application-focused system. Complementing these efforts in ion trap integration, there is parallel progress in developing high-stability, chip-scale lasers for integrated, high-performance light sources \cite{Loh:2025} and integrated detectors for state readout \cite{setzer:2021,reens:2022}. 

In this work, we bridge the application gap by demonstrating an autonomously operating optical clock using an ensemble of up to four \ybion ions in a room-temperature, multi-site, integrated photonics trap. All stages of the clock sequence --- state preparation, coherent qubit manipulation, and detection --- are performed using light delivered entirely by on-chip waveguides. The significance of this achievement lies not in the raw clock performance but in the successful system-level integration of all operational stages, from automated ion loading and shuttling to coherent, interleaved clock interrogation. The realization of this system required overcoming significant practical challenges, including managing on-chip waveguide scattering and optimizing ion placement to balance performance trade-offs across multiple sites. This work's demonstration of sustained operation despite ion loss moves beyond component-level functionality and establishes a manufacturable and resilient architecture for the next generation of portable, scalable, multi-ion quantum sensors and quantum computing systems.

\section{Results}


\subsection{Experimental setup} 
\label{sec:expSetUp} 

\begin{figure*}[!tp]
    \includegraphics[width=\linewidth]{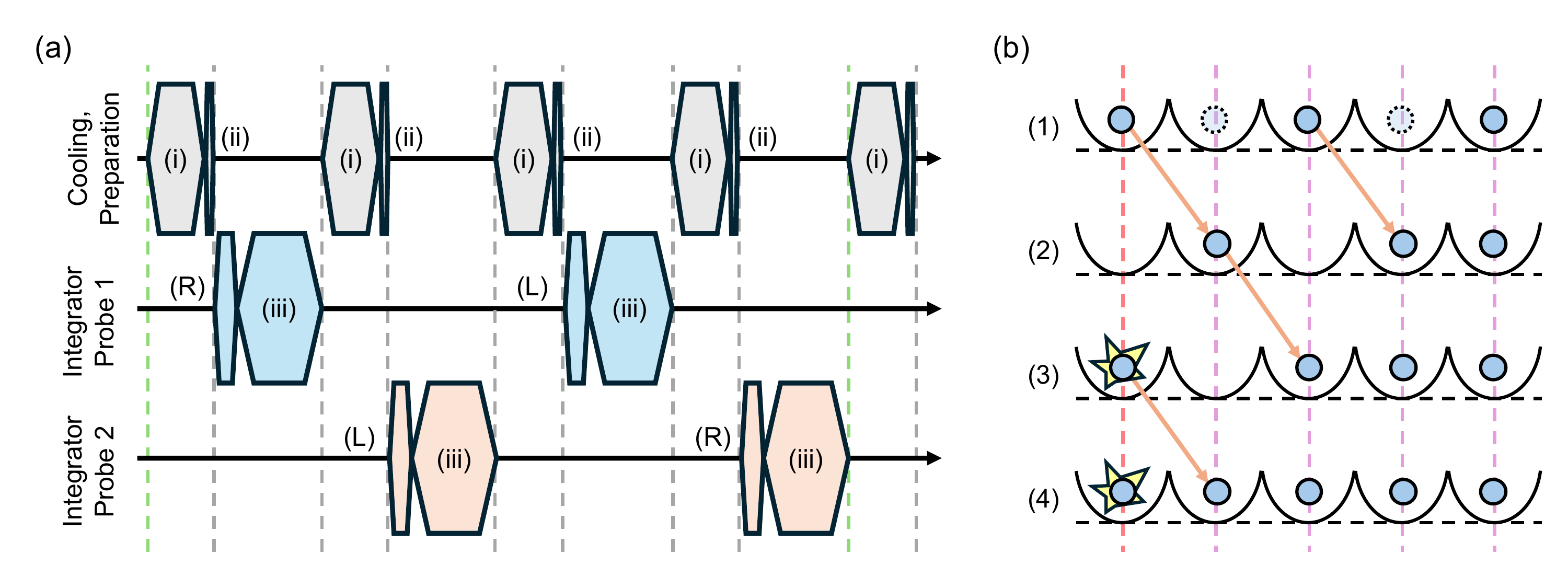}
    \caption{\textbf{Multi-ion clock protocol} \textbf{(a)} Pulse sequence used in the interleaved measurement. The sequence involves (i) cooling immediately followed by state preparation (ii). Afterwards, the sequence is handed off to one of two independent second-order integrators. They apply their followed frequency plus (in the R case) or minus (in the L case) half the transition's spectroscopic linewidth, obtained prior to running the clock. Operation (iii) indicates the read out period of the ion. The green vertical lines indicate when the algorithm updates the center frequency of each integrator. Requests for new ions are also sent at this point if they occur at the reporting period. \textbf{(b)} Cartoon of the refill procedure that occurs as the clock is running. Ions are loaded at a loading site (red, dashed vertical line) and shuttled to clock sites (purple, dashed vertical lines). The star-burst indicates the loading of a new ion. When multiple ions are lost simultaneously, the farthest sites from the loading site are preferentially filled first.}
    \label{fig:SystemSequence}
\end{figure*}

In previous demonstrations of integrated photonic ion traps, we utilized a free-space grating input coupling scheme \cite{ivory:2021,Kwon:2023}. This presented challenges for both coupling efficiency and stability. Our current room-temperature system makes progress in both challenges by directly coupling an array of fibers to the trap. Single-mode, polarization-maintaining fibers in Pyrex v-groove arrays (VGAs) \cite{Uddin:2009} were aligned with a matching array of tapered waveguides at the edge of the trap chip. The input for each beam is split on-trap via multi-mode interferometer (MMI) splitters into five waveguides, which then go to separate clock sites. Each site is addressable with four waveguide output grating windows that deliver light at 369 nm, 435 nm, 935 nm, and 760 nm. Fig.\ref{fig:SystemIntro}(a) shows an illustration of the site layout on the chip.
The optical loss in each waveguide path is primarily due to imperfect input coupling, material absorption, scattering during propagation, and diffraction inefficiency. The total loss for the clock beam (435 nm) exhibits a total loss of -27 dB.
Although it is difficult to identify the contribution of each loss mechanism, measurements from the loop-back optimization suggest that most of the loss is due to the fiber-to-trap coupling.

For this work, we used the 435 nm E2 (clock) transition of \ybion. This additionally requires 369 nm for Doppler cooling and detection and 935 nm for repump and state preparation. A 760 nm beam can be used to pump the $^{2}F_{7/2}$ state to $^{1}[3/2]_{3/2}$ to rejoin the Doppler cooling cycle \cite{PhysRevA104L010802}, however, in our systems it does not significantly improve the overall lifetime of the ions \cite{Kwon:2023}.

New ions are loaded from sublimated \yb vapor produced by an oven that is powered on and off during clock operation. Resonant 399 nm and nonresonant 393 nm beams photoionize a fraction of the vapor, which can then be trapped in the RF and DC confining fields. Due to the power requirements of the 393 beam and the losses from the fiber-to-trap coupling, a free-space beam launch that also included the Doppler and repump beam was used for loading. 
As a proof-of-principle demonstration, loading via waveguide-delivered beams was demonstrated on separate occasions, still with free-space 393 nm and 399 nm beams. All clock algorithm lasers were delivered to the ions via the integrated waveguides and grating windows.

All DC electrodes are individually controlled. This means that the trapping potential of an individual ion at each site can be compensated with minimal effect on a neighboring site. When an ion is loaded, it is quickly transported to a vacant clock site. The high statistical readout contrast necessary for the clock operation depends on the overlap of waveguide emitted beams and RF null. Since each beam has its own axis of emission, the beam overlap is determined by the design and lithography of the diffraction gratings, as are the individual beam intensity distributions. Due to small imperfections in the beam overlap, there was a compromise between fluorescent counts (requiring good 369- and 935 nm beam overlap), the E2 transition Rabi frequency, and the coherence time (efficient cooling versus excess heating). The DC electrodes were tuned to position the ions at each clock site to qualitatively achieve similar values for these quantities across all sites, and to place the ions at the RF null. The axial trapping frequency defined by the DC potential was 1.02 MHz. The radial trapping frequencies defined by the RF were 3.06 and 3.11 MHz. The DC voltages applied to maintain this compromise did not vary from day to day. However, there were variations correlated with the applied 369 nm beam power. It is hypothesized that this is due to the charging of the waveguide output grating windows. When 369 nm power was altered, the correction voltages would often take a few hours to settle. If the 369 nm power was returned to nominal values, the voltage corrections would also return to the nominal values on a similar timescale. We expect to eliminate this dependence in future traps by using conductive coatings on the output grating windows.

\subsection{Clock protocol for multi-ion ensemble operation}

Fig.\ref{fig:SystemSequence}(a) shows the clock pulse sequence used for interrogation. Clock ions are cooled and optically pumped. Then, the ensemble of ions is probed on the E2 transition at half its spectroscopic width on either side of the inferred center for two separate second-order integrators. Probing at the half-width ensures that fluorescence counts are on average present after sufficiently many interrogations if an ion is present. Additionally, this choice gives a large derivative in the count imbalance with respect to frequency error, the basis of the error signal. The count imbalance feeds into its respective integrator to maintain reference to the ion. Each integrator is independent of the other, probing the same transition on the same set of ions. Integrators follow closely as the ions are lost and reloaded at clock sites. For a single integrator, its feedback onto the local oscillator frequency is as follows. The ion is probed on the right and left sides of the transition one half-width from the inferred resonance. If the ion was shelved in the clock state and therefore below the threshold for detection, the value of the interrogation of that side ($S$) is $n_{S} = 0$ ($S$ is either $R$ or $L$). If the ion was not shelved in the clock state, visible to the detection cycle, and therefore above the threshold for detection,  the value of the interrogation of that side is $n_{S} = 1$. The second-order integrator ($I^{(2)}$) is increased by the value of the first-order integrator ($I^{(1)}$). The imbalance ($n_{R} - n_{L}$) is then added to the first-order of the integrator. Together, these steps are as follows:
\begin{equation}
\begin{split}
    I_{i}^{(2)} \mathrel{+}&= I_{i}^{(1)}\\
    I_{i}^{(1)} \mathrel{+}&= n_{i,R}-n_{i,L}
\end{split}
\end{equation}
The total frequency displacement ($\Delta f$) to track the optical resonance of the clock transition is:
\begin{equation}
    \Delta f_{i} = (g^{(1)} I_{i}^{(1)}+g^{(2)}I_{i}^{(2)})\times \text{FWHM},
    \label{eq:freq}
\end{equation}
where $g^{(1)}$ is a first-order gain ($\sim10^{-3}$) and $g^{(2)}$ is a second-order gain ($\sim10^{-6}$). $\text{FWHM}$ is the spectral full-width half-maximum measured before starting the clock algorithm. In our experiment, this is very close to the Fourier-transform-limited $\text{FWHM}$. The frequency displacement and the half-width detuning is then applied by a double-pass acousto-optic modulator (AOM) to the local oscillator light for the next interrogation. For interleaved comparison two different integrators track independent interrogations, resulting in separate inferred frequency displacements. The frequency displacements are used to determine the clock frequency instability in the next section by analyzing the frequency displacement difference $\Delta f_{1} - \Delta f_{2}$.

In addition to managing the integrator feedback, the clock algorithm uses the same measurement to track which ions are still present and whether an ion needs to be reloaded. To determine whether the ion is present, the sum of side interrogations is tracked. Both ions that are outside of the cooling cycle and physically lost from the trap are perceived as lost when the rolling average falls below a threshold. When an ion in a clock-zone is lost, this is reported to the shuttling software. This report occurs every 20 clock cycles. Each clock cycle involves four interrogations: one on each side of the resonance for each integrator. One clock cycle is the distance between the dashed green lines in Fig.\ref{fig:SystemSequence}(a). The reporting rate is limited by the rate at which the receiving software can communicate consistently. The remaining ions are shuttled into vacancies, automatically filling the farthest sites from the loading site first. The new ion at the loading site is stored as a buffer to quickly replenish empty clock sites. An illustration of this can be seen in Fig.\ref{fig:SystemSequence}(b). The lifetime of ions at an individual clock site was correlated with the distance from the loading site. This motivated filling the farthest sites first. One hypothesis for the cause of this effect is that the localized pressure associated with neutral flux and heated surfaces increases collisions, reducing neighboring clock site lifetimes. The ion lifetimes were approximately one minute during clock operation and 15 minutes when individual ions were only being trapped and cooled. Feedback onto the local oscillator occurs as long as any ions remain. Despite this short lifetime, we were able to keep the clock sites fully occupied with four ions more than 45\% of the time. Furthermore, we operated the clock system continuously for more than two hours on multiple occasions, predominantly limited by algorithmic interlocks to abort the clock in cases where the algorithm was unable to determine if a laser had become unlocked, such as no ions in any sites for an extended period of time. This attests to the effectiveness of the automated ion reloading, shuttling, and clock-cycle reintegration capabilities of the system.

\begin{figure}[!tp]
    \includegraphics[width=\linewidth]{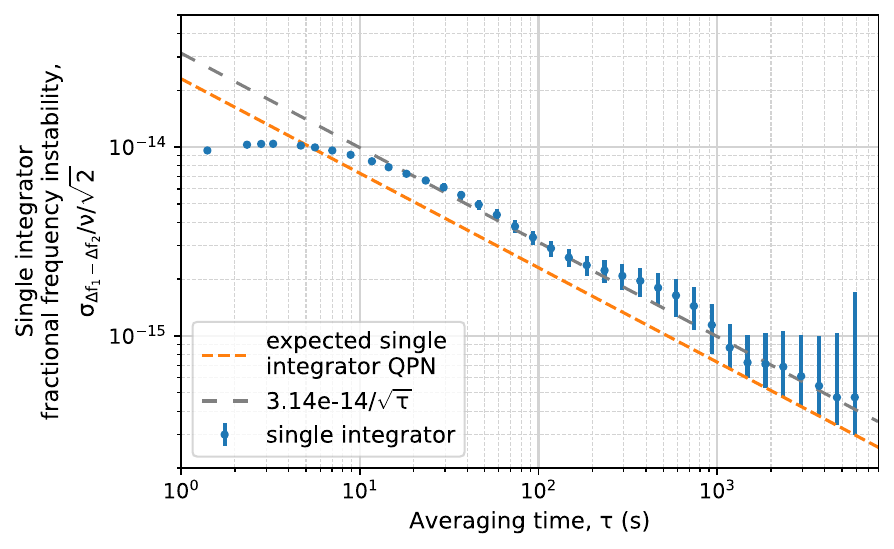}
    \caption{\textbf{Allan deviation.} Single-integrator fractional frequency instability. The fit line of 3.14(5)$\times 10^{-14} / \sqrt{\tau}$ is plotted in gray. The expected quantum projection noise of 4 ions and a $T_{\text{cycle}}$ of four side interrogations is plotted in orange.}
    \label{fig:allanDeviation}
\end{figure}

\subsection{Frequency stability of integrated photonics multi-ion clock}

\begin{figure*}[!tp]
    \includegraphics[width=1.0\linewidth]{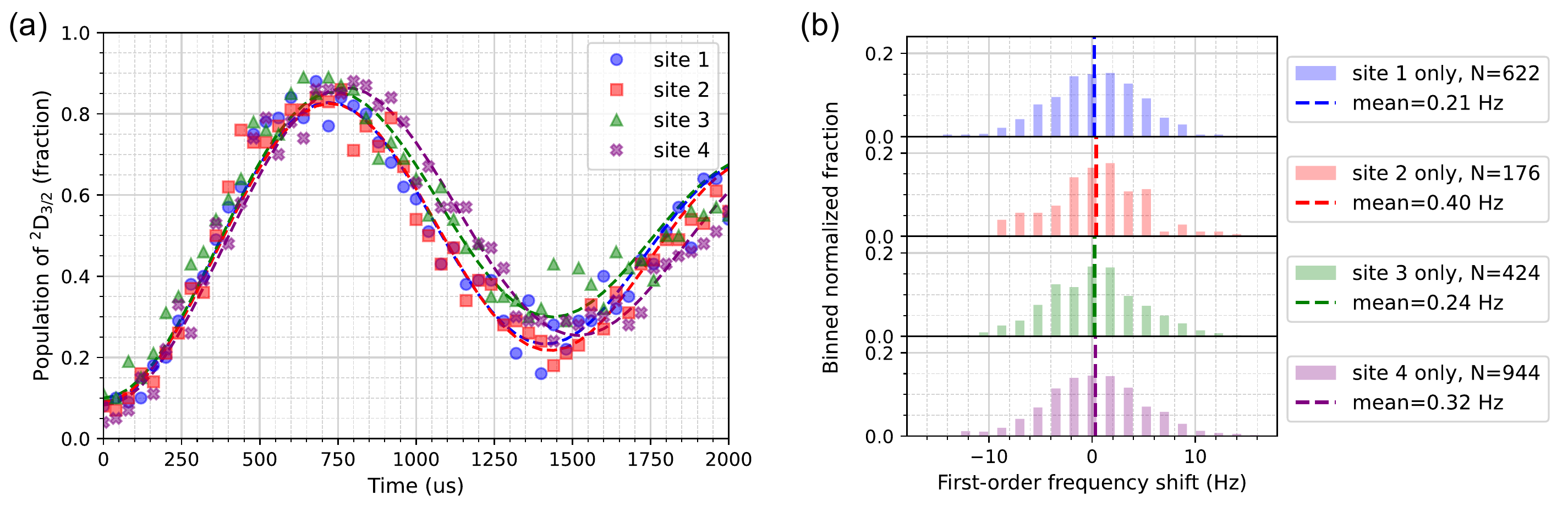}
    \caption{ \textbf{Uniformity across multiple sites.} \textbf{(a)} Representative Rabi flops. Ion positions in the waveguide beams are qualitatively tuned using DC electrodes to achieve Rabi flops with similar features. Fit parameters according to equation (\ref{eq:dephaseFit}) are listed in Table \ref{tab:rabiFit}. \textbf{(b)} First-order frequency shifts for single-ion data. Each mean is consistent with zero.}
    \label{fig:SystemSystematics}
\end{figure*}

Fig.\ref{fig:allanDeviation} shows the performance of the integrated photonics 4-ion ensemble clock. The expected quantum projection noise (QPN) of a single integrator is calculated as follows \cite{PhysRevX14011006}:
\begin{equation}
    \sigma_{\text{QPN}}(\tau) = \frac{1}{2 \pi \nu C T_{\text{probe}}}\sqrt{\frac{T_{\text{cycle}}}{N\tau}},
    \label{eqn:QPN}
\end{equation}
where $\nu$ is the transition frequency, $C$ is the spectroscopic contrast, $T_{\text{probe}}$ is the time the transition frequency is measured, $T_{\text{cycle}}$ is the total cycle time, $N$ is the number of ions probed, and $\tau$ is the averaging time. The expected frequency difference quantum projection noise assumes that all ions are always present ($N=4$). In addition, two integrators interrogate the right and left sides of the resonance. The interrogation on one side of the resonance includes cooling, state preparation, probe, and detection and takes time $T_{S}$. Thus, the total cycle time ($T_{\text{cycle}}$) for the interleaved measurement is four side interrogations (the distance between the dashed green lines in Fig.\ref{fig:SystemSequence}(a), $T_{\text{cycle}} = 4T_{\text{S}}$). Finally, the quantum projection noise of the individual integrators sums in quadrature for the interleaved frequency differences ($\sigma_{\Delta f_{1}-\Delta f_{2}}^{2} = \sigma^{2}_{\Delta f_{1}} + \sigma^{2}_{\Delta f_{2}}$). Assigning equal weight to both integrators to extract the single-integrator value, we report a single-integrator clock with a short-term fractional frequency instability of 3.14(5)$\times 10^{-14} / \sqrt{\tau}$ and expected quantum projection noise instability ($T_{\text{cycle}} = 4T_{S}$, $N=4$) of 2.30$\times 10^{-14} / \sqrt{\tau}$ for an individual clock.

These comparisons demonstrate that the clock performs within a factor of two of the expected quantum projection noise. The deviation is likely caused by technical noise sources. Ion number fluctuations may account for up to 25\% of the fractional frequency instability for the data in Fig.\ref{fig:allanDeviation} based on the weighted average of the quantum projection noise for the observed ion occupancy. However, this still leaves some instability above the quantum projection noise unaccounted for. If we were to operate a clock with all interrogations used to update a single integrator, the update cycle would be improved by a factor of two, $T_{\text{cycle}}=2T_{S}$. The resulting clock would have an expected fractional frequency instability of 2.2$\times 10^{-14} / \sqrt{\tau}$ and a quantum projection noise instability ($T_{\text{cycle}} = 2T_{S}$, $N=4$) of 1.6$ \times 10^{-14} / \sqrt{\tau}$. We were unable to directly measure the single-integrator clock stability because we did not have access to a frequency reference with sufficient stability.

\subsection{Characterization of multi-site uniformity for future scalability}

Site uniformity is of great interest to scalable designs for atomic clocks, quantum sensors, and quantum computation, more broadly. Small shifts in the clock transition can lead to absolute frequency and gate errors. Efforts to calibrate out site-specific differences work on a small scale, but the number of calibrations becomes more tedious and the maintenance becomes intractable as the number of sites grows. Therefore, it is desirable to achieve general uniformity across all sites.

The direction and focal size of each waveguide output beam is lithographically defined. Before running the clock algorithm, the ions were verified to have similar counts, Rabi frequencies, and coherence times. The electrodes must be finely tuned to match these, as small local differences are hard to predict. We also observed a dependence of the compensation voltages on the 369 nm beam power. Because the timescale of the effect was hours, the small periods of the 369 nm beam being off for the clock transition excitation were inconsequential. Once established, the compensations would hold for several weeks. While we expect to eliminate this dependence in future traps by using conductive coatings on the output grating windows, the power drift was not an issue unless the laser power was changed by roughly a factor of 2 or more. Fig.\ref{fig:SystemSystematics}(a) shows a representative set of Rabi flops from an ion at each site. Measurements are modeled similarly to \cite{semenin2022determination}. The ions are trapped in separate DC potentials 180\mic apart, so the chain configuration terms can be dropped, resulting in the following equation for the probability of each ion being shelved in the clock state:
\begin{equation}
    \begin{split}
    &P(t) = c+\frac{A}{2}\\
    &\times\left[1-\text{Re}\left(\frac{e^{i \Omega_{0} t}e^{-i \Omega_{0} \eta^{2}t/2}(1- \frac{\bar{n}e^{i\Omega_{0}\eta^{2}t}}{\bar{n}+1})}{(\bar{n}+1)-2\bar{n}\cos(\Omega_{0}\eta^{2}t)+\frac{\bar{n}}{\bar{n}+1}}\right)\right],
     \end{split}
    \label{eq:dephaseFit}
\end{equation}
where $A$ and $c$ determine our spectroscopic contrast, $\Omega_{0}$ is the Rabi frequency, $\eta = \frac{1}{\lambda}\sqrt{\frac{h}{2mf_{\text{sec}}}}$, and $\bar{n}$ is the average occupation of the corresponding secular motion. $\eta$ is defined prior to fitting by choice of secular frequency ($f_{\text{sec}}$), species mass ($m$), and transition wavelength ($\lambda$). The bias is added to the fit to account for stray photons that scatter into the PMT. The fit values are shown in Table \ref{tab:rabiFit}.

During the clock protocol, and as part of the shuttling procedure, the algorithm reports which ions are present. Additionally, the algorithm reports the summations of the interrogations for each side in a reporting period $j$ as the respective sum, $\left( s_{j,S} = \sum _{rj}^{r(j+1)-1} n_{i,S} \right)$. A reporting period ($r$) is 20 clock cycles for this experiment. Post-processing of these data allows the extraction of information about site-specific shifts. To analyze the site-specific shift, we define a related quantity, the first-order frequency shift:
\begin{equation}
    \begin{split}
    \Delta f_{j}^{(1)} = & g^{(1)}(s_{j,R} - s_{j,L}) \times \text{FWHM}
    \end{split}
\end{equation}
This information is then sorted by which ions are present. Fig.\ref{fig:SystemSystematics}(b) shows the first-order frequency shift constrained to only one ion at a specific site. These data are sparse but are most sensitive to individual sites. Due to the ion population statistics, the points are also predominately preceded by reporting periods containing two or more ions. Based on this analysis, if a site-specific shift exists, it is below the first-order servo error of our system (1.75 Hz) and statistically consistent with no shift between sites during the experiment.
\begin{table}[!htp]
    \centering
    \begin{tabular}{c|c c c c}
        site & $A$ & $\Omega_{0}$ $(2\pi\text{Hz})$ & $\bar{n}$ & c\\
        \hline \hline
         1 &  0.80(4) & 5002(59) & 28(3)& 0.09(2)\\
         2 &  0.78(4) & 4920(65) & 25(3)& 0.09(2)\\
         3 &  0.85(5) & 5005(65) & 37(4)& 0.10(2)\\
         4 &  0.86(3) & 4689(46) & 31(3)& 0.08(2) \\
    \end{tabular}
    \caption{Fit parameters for the qualitatively-tuned to match, representative Rabi flops in Fig.\ref{fig:SystemSystematics}(a) according to equation (\ref{eq:dephaseFit}). The secular trapping frequencies for this experiment were 1.02 MHz, 3.06 MHz, and 3.11 MHz. The lower axial frequency was used for the secular frequency in the defined value of $\eta$ to form a conservative estimate of $\bar{n}$, despite the laser projecting partially onto all motional modes.}
    \label{tab:rabiFit}
\end{table}

\section{Conclusion} 

In this work, we present an autonomously operating, trapped-ion optical clock built on a multi-site integrated photonics chip with an ensemble of $^{171}$Yb$^{+}$ ions. The fractional frequency instability of the resulting clock is 3.14(5)$\times 10^{-14} / \sqrt{\tau}$. We also demonstrate that individual sites do not exhibit clock transition frequency differences greater than the first-order servo error of 1.75 Hz. This marks a milestone for the realization of scalable photonic ion traps for clocks, sensors, and quantum computation.

Near-term improvements can be achieved by improving the vacuum pressure, which has gradually degraded as a result of tiny leaks in the fiber feedthroughs and out-gassing from the epoxy coated acrylite fibers. These materials do not meet the highly restrictive out-gassing standards required to achieve significantly longer lifetimes. With a better choice of UHV materials, an improved vacuum will extend the ion lifetime by preventing collision-induced loss and micro-motion. The reduced collisional decoherence should also allow us to further lower the quantum projection noise by using longer Ramsey sequences for interrogation instead of Rabi pulses.

This trap suffers from high heating rates, as discussed in more detail in the Methods. Although this was not likely the limiting factor in clock performance, future systems with improved detection efficiency and ion lifetime would benefit from cooler atoms and the associated improved contrast and coherence time. One culprit for high heating rates could be noise due to the buildup of charge on the dielectric surfaces of the grating windows \cite{PhysRevLett126230505}. This could be alleviated by the inclusion of a conductive but transparent coating over the grating windows, such as indium-tin-oxide. In addition, side-band cooling was not implemented. Side-band cooling was time prohibitive, in part because of the low optical coupling from fiber-to-trap. Better optical coupling would allow for side-band cooling and increase both contrast and coherence time.

The ion lifetime in this work was not optimal (on the order of a minute) and is generally considered to be too short to operate as an optical clock. However, one of the strengths of this system is highlighted by overcoming this hurdle by automatically loading and shuttling the ions while simultaneously performing interrogations. The fact that no lapse in clock operation occurred is critical and is directly beneficial for various applications. Moreover, since the protocol used in this work can be scaled to use many ions arranged in multiple ensembles, we believe that this approach using high levels of integration at the chip level can support both more accurate and more deployable atomic clocks.

\subparagraph{\textbf{Acknowledgments.}}
Sandia National Laboratories is a multi-mission laboratory managed and operated by National Technology \& Engineering Solutions of Sandia, LLC (NTESS), a wholly owned subsidiary of Honeywell International Inc., for the U.S. Department of Energy’s National Nuclear Security Administration (DOE/NNSA) under contract DE-NA0003525. This written work is authored by an employee of NTESS. The employee, not NTESS, owns the right, title, and interest in and to the written work and is responsible for its contents. Any subjective views or opinions that might be expressed in the written work do not necessarily represent the views of the U.S. Government. The publisher acknowledges that the U.S. Government retains a non-exclusive, paid-up, irrevocable, world-wide license to publish or reproduce the published form of this written work or allow others to do so, for U.S. Government purposes. The DOE will provide public access to results of federally sponsored research in accordance with the DOE Public Access Plan.

This work was supported in part by the Laboratory Directed Research and Development program at Sandia National Laboratories; the U.S. Department of Energy Office of Science National Quantum Information Science Research Centers, Quantum Systems Accelerator; and the Defense Advanced Research Projects Activity (DARPA).

D.R.L. was supported in part by the National Science Foundation Q-SEnSE Quantum Leap Challenge Institute (Grant Number OMA-2016244).

\subparagraph{\textbf{Data availability}}
All relevant data are available from the corresponding authors upon reasonable request.

\newpage
\label{sec:methods}

\subparagraph{\textbf{\centerline{Appendix A: Experimental system}}}

\subparagraph{\textit{Trap design, fabrication and vacuum}}
The experiment was conducted using a trap designed and fabricated at Sandia National Laboratories. Although this trap shares features with previous designs \cite{ivory:2021,Kwon:2023}, it uniquely incorporates a 1$\times$5 multi-mode interferometer (MMI) waveguide splitter for simultaneous multi-ion clock interrogation. \ybion is a good species for deployable and robust systems due to its suppressed first-order Zeeman shift. The large wavelength range (369 nm to 935 nm) adds to the design complexity, as shorter wavelengths require unique waveguide designs and are more sensitive to fabrication tolerances and feature roughness. All waveguides, including 369nm and 435nm, were made with silicon nitride.  This is made possible using a plasma-enhanced chemical vapor deposition (PECVD) nitride film which Sandia National Laboratories has optimized for low propagation loss at blue and UV wavelengths.

Fiber coupling was accomplished using a UHV vacuum-compatible glue attachments by Optelligent of v-groove arrays (VGA). Single-mode polarization-maintaining fibers of the appropriate core diameters (to ensure single-mode propagation at the appropriate wavelength) were placed in Pyrex v-groove arrays prior to attaching [31]. These were then aligned to a matching array of tapered waveguides on the edge of the trap chip. These tapered waveguide array edge-couplers are located transverse to the trap axis to minimize the scattered light from the coupling region into the readout optics. The attachment was optimized by maximizing the light that circulates through an additional on-chip loop-back waveguide. For a fiber feedthrough, a 2.75-inch CF flange was drilled with diameters of the holes large enough to fit several fibers. The fibers were inserted and epoxied in place using low out-gassing 3M Scotch-Weld 2216 epoxy.

The on-trap MMI splits a single input for each beam into five waveguides, which go to separate clock sites 180\mic apart. Although there are five sites per splitter, only four are used for clock operation. The specific trap used in this system had several malfunctioning electrodes at the fifth site, making that site functionally inoperable for tuning. Due to the same electrodes, shuttling past this site to access a farther ensemble of clock sites was frequently unsuccessful and deemed not useful in a demonstration.

\subparagraph{\textit{Ion trapping and shuttling}}
Vaporized ytterbium atoms are provided through a loading slot, roughly 2\mic $\times$ 22\mic, which penetrates through the trap electrodes and substrate. By applying a current to an oven containing ytterbium metal located a few centimeters below the loading site, a sufficient neutral vapor for loading is generated. A two-photon process using 399 nm (resonant) and 393 nm (non-resonant) lasers photoionizes the neutral ytterbium. It is important to note that these beams are only used for ionization and not for operations involving the clock algorithm. Due to its non-resonant nature and high power requirement, the 393 nm beam was delivered via a free-space launch; the 399 nm beam was combined with it for simplicity. The necessary Doppler and repump beams for the loading site were also delivered via free-space, which proved helpful for diagnostic purposes. The loading site also has its own set of waveguides (without an MMI splitter), which require approximately five times less power to operate than the clock site waveguides. Waveguide-based loading was demonstrated on a separate occasion. In both cases, the ions typically load in 10s of seconds.

The DC and RF fields provided by the trap electrodes confine the newly ionized atom at the loading site when laser cooled. The RF field is driven at approximately 40 MHz to confine the ion radially with trapping frequencies near 3.1 MHz. DC fields confine ions axially with a trapping frequency of 1.02 MHz. RF power was monitored by a Texas Instruments LMH2120 power detector chip with a 1\% capacitive pick-off circuit between the helical resonator and the trap RF input and is used in a servo to maintain constant RF power. The trapping pseudo-potential defined by the RF field holds ions approximately 50\mic above the electrode surface, where all waveguide beams are expected to converge at their focuses at the interrogation sites. We measure an average motional quanta of $\bar{n} \sim 30$ quanta based on the fitting of Rabi oscillations \cite{semenin2022determination} with a detection fidelity of 87\%, which is comparable to previous research \cite{Kwon:2023}. This low detection fidelity is due to the low numerical aperture and relatively high background scattering rate inherent in side-imaging.

We note that the heating rate of the system is high compared to our previous work \cite{Kwon:2023}. We think this is due to a combination of factors. For one, the ion could not be placed simultaneously on the RF-null at an optimal position for Doppler cooling while maintaining good overlap with the clock laser waveguide output. As a result of the ion being off the RF null, it is more susceptible to amplifier noise in the trap RF. We believe that this is due to imperfect trap fabrication.

Additional heating could also be due to the charging of the waveguides by the 369 nm beam. As stated in previous sections, we noticed a correlation between the input 369 nm laser power and compensation values of the DC electrodes used to position the ions in the 369 nm and 435 nm beams. The charging effect also suggests that the ions have additional exposure to the dielectric surfaces and an implied increase in Johnson noise heating. This high heating rate prevents us from obtaining higher clock resolution with the Ramsey method and remains an improvement for future work.
\begin{figure}[!tp]
    \includegraphics[width=1.0\linewidth]{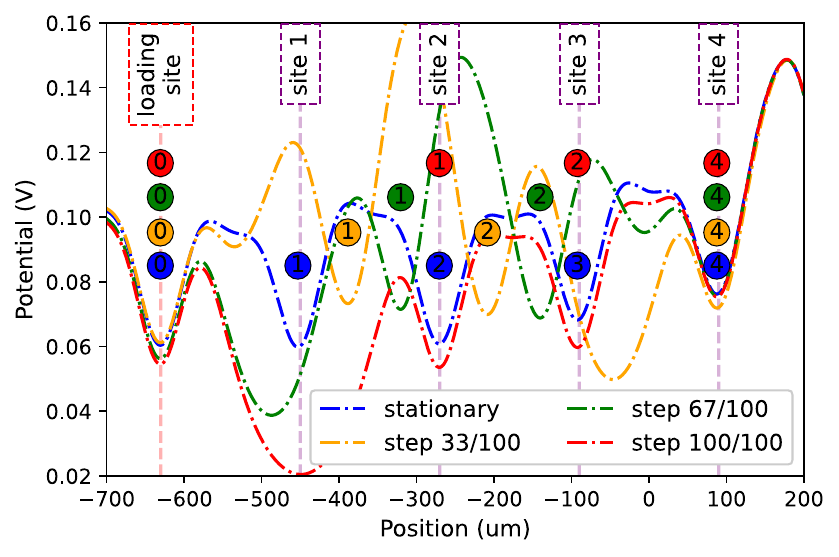}
    \caption{A sample of the DC potential at the ions' height above the chip. Here ions are shuffled from sites 1 and 2 to sites 2 and 3, respectively, to fill a vacancy at site 3. Ions at the loading site and site 4 remain stationary and have minimal perturbations to their respective DC axial trapping potentials. The minimums containing ions are labeled according to the site they originated from in "stationary" case. The colors indicate the step to which they belong. Vertical dashed-lines mark the regions addressable by different waveguides. The loading region marked with a red vertical dashed-line. The clock sites used in this experiment are marked with purple vertical dashed-lines.}        
    \label{fig:voltageExample}
\end{figure}

The trapped ions are moved to their target locations within a site by adjusting the DC potentials. Each location needs a specific field profile due to small fabrication errors and charging effects. Most importantly, these profiles are tuned to achieve identical Rabi dephasing and $\pi$/2 pulse times. Therefore, we first used a dynamic solution to obtain precise control \cite{Kwon:2023} and to optimize the compensation map for the trap. The dynamic solver provides a real-time calculation of the potential, which allows for easy compensation adjustment for various conditions. The dynamic solver allows for adjustment of a single trapping potential while maintaining the compensation fields at other trapping sites. For the actual clock algorithm, the DC voltages are compiled into a discrete set to reduce computational load and to increase and regularize the shuttling speed. This is critical for maintaining multi-ion occupancy for several hours. Consistent, fast shuttling reduces the accumulated motional quanta, which is important for high heating rate traps. The loss of an ion during shuttling becomes even more important as the number of ions increases. The ability to replenish lost ions depends on the mean loading rate being greater than the ion loss rate. The ion loss rate should scale as $N \Gamma$ for $N$ ions and a single ion loss rate of $\Gamma$. We addressed this by categorizing each ion loss event and mapping out a shuttling process that keeps the remaining ion sites stable. Additionally, up to half of the ensemble ions are shuttled simultaneously to minimize the number of required shuttling steps while not blocking the probe of clock ions. For the specific case described in this paper, four different shuttling scenarios cover the replacement of ions at the clock sites. All scenarios begin and end with all potential wells present at all sites. Fig.\ref{fig:voltageExample} shows an example of some steps in the compiled solution used in a single scenario. Step 0 is the state of all wells present at all sites. Each shuttling from one site to the other site takes less than a second. The number of shuttling scenarios is proportional to the number of ions, making this a conceivable path forward to managing large numbers of ions. The remaining limitations are the ion lifetimes, the tuning of individual sites, and the evolution of the stray field. These could be incorporated as part of a tuning procedure or possibly even dynamically adjusted for in more sophisticated algorithms. However, our stray field evolution did not occur on any meaningful timescale as long as 369 nm beam power remained nominal.

\subparagraph{\textit{Side-imaging and signal optimization}}
We applied a similar side-imaging technique as in our previous research \cite{Kwon:2023}, but in a modified way. A 2-inch objective lens with a numerical aperture (NA) of 0.12 was placed at the side window to exclusively capture maximum ion fluorescence while minimizing light noise from waveguide emissions. We placed extra pinholes of 100\mic diameter in front of the 1D array PMT (Hamamatsu H11659-200). Each pinhole was separated by 1 mm, equal to the pitch of the 1D PMT array. The image magnification was coarsely tuned by obtaining an image from a camera (Qimaging ROLERA em-c$^2$). By coupling light to the integrated waveguides via fiber edge coupling, the stray light at the detectors was significantly decreased. With optimal alignment, we achieved a signal of $>2.5$ kcps with $\approx 0.1$ kcps of noise, an SNR greater than 25. This is about five times better than the SNR from our previous work, allowing us to decrease the detection time to 3 ms.

\subparagraph{\textbf{\centerline{Appendix B: Clock interrogation}}}
\begin{figure}[!tp]
    \includegraphics[width=\linewidth]{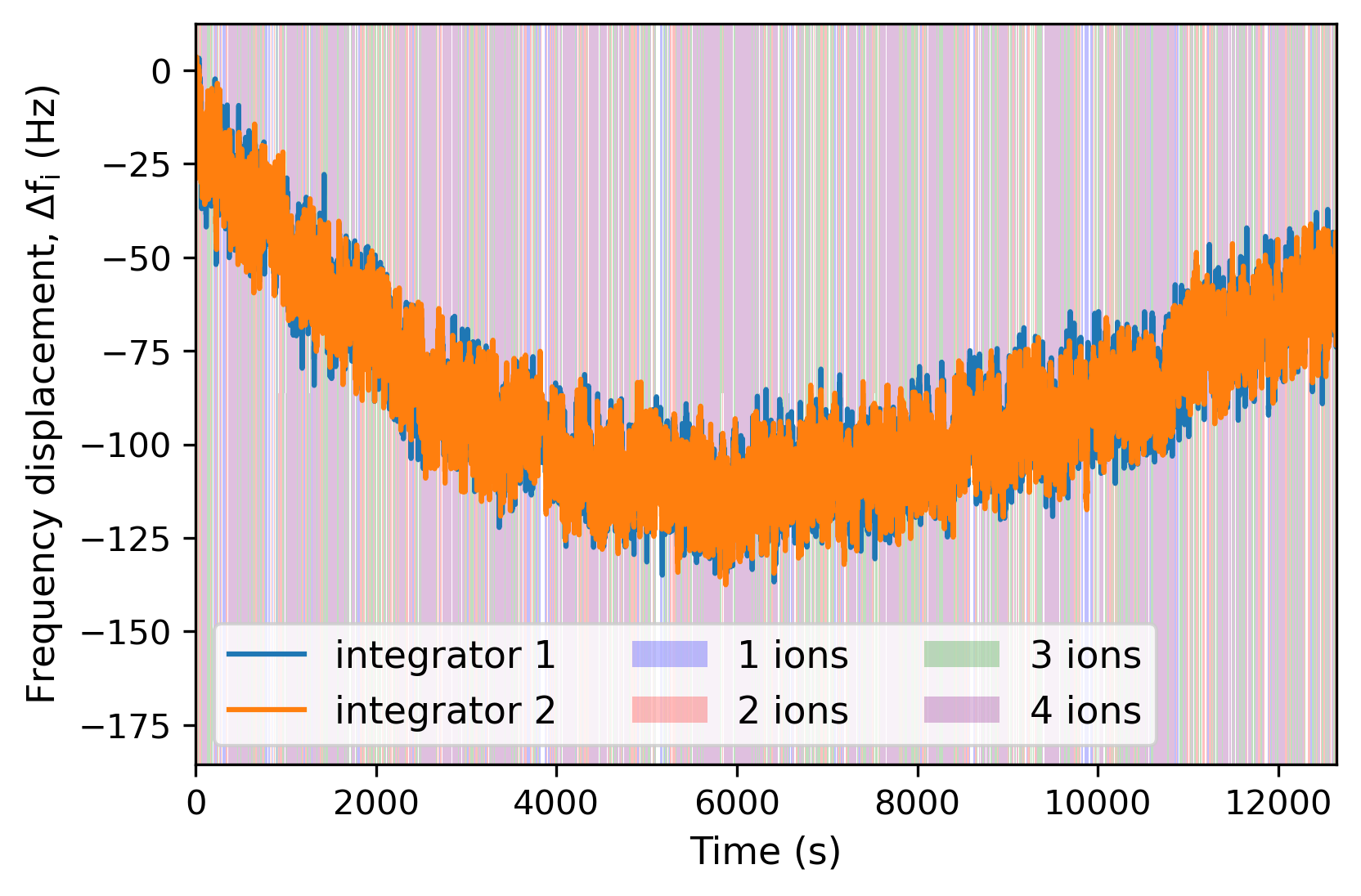}
    \caption{An example of the two second-order integrators as they track the same ions. Colors above and below the mid-line indicate how many ions the algorithm claims to have. Each integrator independently tracks the ions present for diagnostics.}
    \label{fig:independentfrequencies}
\end{figure}

\subparagraph{\textit{Gain and feedback}}
In a given clock cycle, the ion is probed at half its spectral linewidth from resonance. in circumstances where the clock frequency is at the correct position, the ion will be shelved in the $^{2}D_{3/2}$ state 50\% of the time for either probe. In the case where the clock frequency is not properly centered, one probe will be more likely to fluoresce in detection. This forms the basis of the error signal for the feedback. Fluorescent counts are compared to a threshold to determine the state of the ion. A second-order with a smaller fraction of the FWHM is incremented by the first-order integrator to balance the first-order integrator. Then feedback is applied as a fraction of the FWHM of the spectral linewidth via a first-order integrator. An example of the frequency displacement of these two second-order integrators as they track the same ions is shown in Fig.\ref{fig:independentfrequencies}. The interplay between the clock cycle length, the FWHM of the spectral line, and the first-order gain is the main contributor to our servo bandwidth of approximately 5 seconds (approximately 800 interrogations).

No additional probes are performed on or off resonance, as the counts accumulated during a probe are sufficient to determine if an ion is present for reloading. An average of the fluorescence is maintained during the clock cycle. When the counts summed from both sides drop below a separate threshold, the ion is treated as lost. This can occur when an ion is actually lost or a collision with background gasses puts the ion in a state inaccessible to our lasers.

\subparagraph{\textit{Clock protocol}}
The ion presence determination and shuttling occur at significantly different timescales. A single clock interrogation occurs in roughly 6 ms. However, the shuttling is controlled by a separate slower system. Additionally, electronics on the receiving end limit reliable communication from the FPGA system that manages the clock algorithm to once every 12 clock cycles. Thus, the clock algorithm reports every 20 clock cycles ($\approx$ 0.47 s) with information about which ions are still present and additional data used in post-processing for diagnostics. This time allows ions to go dark and subsequently return to the cooling cycle by collision if they go dark. If the ion is dark outside of the cooling cycle for a sufficient number of clock cycles, the ion is treated as lost and will be marked by the algorithm for reloading.

\bibliographystyle{apsrev}
\bibliography{clockRef_arxiv}

\end{document}